\documentclass[twocolumn,showpacs,showkeys,prd]{revtex4-1} 
\usepackage{color}
\usepackage{ulem}
\usepackage{epsfig,dsfont,amssymb,amsmath,amsfonts,amsbsy,mathrsfs,hyperref}
\usepackage{amsmath,bm}
\usepackage{graphicx}
\usepackage{amsmath}
\usepackage{amssymb}
\usepackage[space]{ctex}
\begin{document}

\title{Magnetic Skyrmions are Quasi-Magnetic Monopoles\\
in Two-Dimensional Magnetic Materials}

\author{Ji-Rong Ren}
\email{renjr@lzu.edu.cn}

\author{Zhi Wang}
\email[Corresponding Author: ]{zhwang17@lzu.edu.cn}

\author{Fei Qu}

\author{Hao Wang}

\affiliation{Institute of Theoretical Physics  \& Key Laboratory for Magnetism and Magnetic Materials of the MOE, Lanzhou University,  P.R.China, 730000}

\begin{abstract}
Using $su(N)$ Cartan subalgebra local bases parametrization of density operator $\rho$, we prove that the Wu-Yang potentials of a general $N$-level quantum system are completely expressed by $SU(N)$ gauge transformation. By taking the $SU(2)$ Cartan subalgebra local basis as a local normalized magnetization vector, we find that magnetic skyrmion and Wu-Yang magnetic monopole have the same algebraic structure. Moreover, by taking the adiabatic unitary evolution of magnetization as local gauge transformation, we verify that the Wu-Yang potential of magnetic skyrmions is proportional to the Berry connection, this means that magnetic skyrmion is quasi-magnetic monopole.  The exact relation between Wu-Yang potentials and Berry connection is discussed in detail for the general $SU(N)$ case, i.e. the Berry connection for the pure or mixed state is the weighted average of the $(N-1)$ Wu-Yang potentials.
\end{abstract}
\maketitle

\section{Introduction}
Magnetic skyrmions are topological spin configurations in magnetic materials, which usually originate from the chiral Dzyaloshinskii-Moriya interactions(DMI)\cite{skyrob1, skyrob2, DM}. Experimentally, there are many methods to create and delete individual magnetic skyrmion at a given position of magnetic materials\cite{skyrapp}, which provide important technical support for the potential application of magnetic skyrmions to information storage and processing\cite{skyrapp, skyappl}. All these methods come from the rotation of the direction of local magnetization but not a displacement of its location\cite{skyrapp, RenWangWangQu, logicgate2015, Nikals1}.
Essentially, in theoretical physics, this kind of rotation of local magnetization can be treated as a local gauge transformation for magnetization\cite{BrunoDugaevTaillefumier,Tatara}.
An important property of magnetic skyrmion is the existence of its emergent gauge field, this field is explained as real space Berry connection which generally is derived from adiabatic approximation\cite{skyremg, emgfield, realBerry}. So our questions are, what exactly is the relation between Berry connection, gauge transformation and adiabatic approximation for $SU(2)$ magnetization in magnetic skyrmions, and whether there exists a general relation for $SU(N)$ case.

We will find that the $SU(N)$ gauge transformations are rotation matrices for the general $N$-level quantum system with any state expressed as density operator $\rho$. Analogous to the 't Hooft $SU(2)$ gauge-invariant electromagnetic tensor of magnetic monopoles\cite{Hooft1974},
using Duan-Ge gauge potential decomposition theory\cite{DuanGe1979}, Duan and Zhang found $(N-1)$ $U(1)$ magnetic monopole gauge fields, i.e. Wu-Yang potentials\cite{Wu-Yang1}, of the extended $SU(N)$ gauge theory\cite{DuanZhang,LiuDuanZhang2005SkySUN,DuanRen}. These Wu-Yang potentials are completely expressed by local gauge transformations $U(x)\in SU(N)$.
We will study the exact relation between Wu-Yang potentials and Berry connection for a general $N$-level quantum system, this is an interesting extension for $SU(2)$ magnetization because the density operator includes the mixed state also.
We develop a Cartan subalgebra local bases parametrization method for the density matrix of a general $N$-level quantum system\cite{mixgeo2}.
We can prove the Wu-Yang potentials of this quantum system are the projection of $su(N)$ flat connection $\frac{1}{ig}\partial_{\mu}UU^{\dag}$ on $(N-1)$ Cartan subalgebra local bases $n_{i}$, which theoretically relate to $(N-1)$ topological quasi-particles similar to magnetic skyrmions in $SU(2)$, this provides us a new clue for creating topological quasi-particles.

In general, despite $U(x)$ adiabatically evolution or not, the $(N-1)$ Wu-Yang potentials always exist and relate to general geometric phase of pure or mixed state\cite{mixgeo}.
However, when the adiabatic approximation is satisfied, by taking the adiabatic unitary evolution as a local gauge transformation $U(x)$, we will find that $N$ Berry connections of the $N$-level quantum system can be expressed by $U(x)$. The weighted average for these Berry connections with respect to the eigenvalues of density operator is equal to the expectation value of flat connection with respect to the final state of the adiabatic unitary evolution. An exact relation between Wu-Yang potentials and the weighted average of Berry connections is obtained in the condition of adiabatic approximation.
We will find in the case of $SU(2)$ magnetic skyrmions, the Berry connection is proportional to Wu-Yang potential exactly, and the magnetic skyrmion and Wu-Yang magnetic monopole have the same algebra structure. Therefore, the magnetic skyrmions are quasi-magnetic monopoles in magnetic materials, which is related to the abundant topological structure in Skyrme theory\cite{Cho1, Cho2}.

This paper is organized as follows. In section II, we study the Wu-Yang potentials of a general $N$-level quantum system, which is based on the parametrization of density operator by $su(N)$ Cartan subalgebra local bases.
In section III, we discuss the charge of Wu-Yang magnetic monopole for $SU(2)$ case, it naturally relates to a magnetic skyrmion charge.
In section IV, we take the $su(2)$ Cartan subalgebra local basis as the local magnetization of magnetic material, so the magnetic skyrmion is analyzed by gauge transformations.
In section V, we study the exact relation between $SU(N)$ Wu-Yang potentials and Berry connection in adiabatic approximation.

\section{The Wu-Yang Potentials of $N$-level quantum system}
In this section, we analyze the emergent gauge fields for a general $N$-level quantum system by using $su(N)$ Cartan subalgebra local bases parametrization.

For a general $N$-level quantum system, its quantum states are described by density matrix $\rho$. The density matrix $\rho$ must satisfy three properties $\rho^{\dag}=\rho$, Tr$\rho=1$ and $\rho\geq0$ , where $\rho\geq0$ is called the positive semidefinite condition of $\rho$, and it means all eigenvalues of $\rho$ are nonnegative. We know any $(N\times N)$ Hermitian matrix can be expanded in $su(N)$ Lie algebra bases $T_{a}$ and the unit matrix $I_{N}$. Considering the normalization property Tr$\rho=1$, the $su(N)$ Lie algebra vector representation of density matrix $\rho$ is\cite{KimuraG, ByrdKhaneja}
\begin{equation}\label{rho}
\rho=\frac{1}{N}(I_{N}+\sqrt{2N(N-1)}v^{a}T_{a}),
\end{equation}
where the Lie algebra bases $T_{a}(a=1,2,\cdots,N^2-1)$ are generators of $SU(N)$ Lie group in fundamental representation and the Lie algebra vector is $v\equiv v^{a}T_{a}\in su(N)$. We use the convention of summation over repeated indices.

The vector space for $su(N)$ Lie algebra is composed of vector bases $T_{a}$, when we use the conventions of normalization relations
\begin{equation}\label{nor}
\textrm{Tr}(T_{a}T_{b})=\frac{1}{2}\delta_{ab}
\end{equation}
and the commutation and anticommutation relations
\begin{equation}\label{com}
[T_{a},T_{b}]=if_{abc}T_{c}, \;\;\;\{T_{a},T_{b}\}=\frac{1}{N}\delta_{ab}I+d_{abc}T_{c},
\end{equation}
the constant in eq.(\ref{rho}) ensure that $(v,v)=1$ for pure states and $(v,v)<1$ for mixed states, where the definition of the inner product of Lie algebra vectors is
\begin{equation}\label{inn}
(T_{a},T_{b})\equiv 2Tr(T_{a}T_{b}).
\end{equation}

The restricted region of $v$ from the positive semidefinite condition of $\rho$ was analytically obtained in ref.\cite{KimuraG, ByrdKhaneja}. More specifically, since the eigenvalues of $\rho$, we denote it as $a^{n}(n=1,2,\cdots,N)$, are invariant under unitary transformations, the necessary and sufficient condition for density matrix $\rho$, as described in ref.\cite{mixgeo2, ByrdBoyaMimsetal}, is
\begin{equation}\label{rhod}
\rho=U\rho_{d}U^{\dag},
\end{equation}
where $\rho_{d}\equiv a^{n}\rho_{n}$. The $\rho_{n}(n=1,2,\cdots,N)$ is the pure state density matrix
\begin{equation}\label{rhon}
\rho_{n}=\textrm{Diag}(0,\cdots,0,1,0,\cdots,0)
\end{equation}
with the single 1 in the $n$-th diagonal. The $a^{n}$ satisfies $\sum_{n}a^{n}=1$ and $0\leq a^{n}\leq1$. In the following, we study the case of non-degenerate for the non-zero $a^{n}$, for convenience, we assume $a^{1}=a^{2}=\dots=a^{m-1}=0(m\in\mathbb{Z},1\leq m\leq N)$ and
\begin{equation}\label{eig}
0<a^{m}<a^{m+1}<\cdots<a^{N}\leq 1.
\end{equation}
The $U$ in eq.(\ref{rhod}) is any special unitary transformation $U\in SU(N)$.

For a given element $U(x)\in SU(N)$, the local bases of $su(N)$ Lie algebra are defined as
\begin{equation}\label{Localbases}
u_{a}(x)=U(x)T_{a}U^{\dag}(x),
\end{equation}
where we use $x=\{x^{\mu}\}(\mu=1,2,\cdots,D)$ to represent the coordinate of the $N$-level quantum system in any $D$-dimensional parameter space, which can be the real space, the momentum space and other parameter spaces, and we assume that the density operator is defined smoothly and single-valued in these parameter spaces.

Moreover, the Cartan subalgebra of $su(N)$ Lie algebra is composed of diagonal generators $H_{i}(i=1,2,\cdots,N-1)$ which satisfy $[H_{i},H_{j}]=0$. In this vector space of $su(N)$ Cartan subalgebra, the local bases are defined as
 \begin{equation}\label{nix}
   n_{i}(x)=U(x)H_{i}U^{\dag}(x).
 \end{equation}
It can be proved that $n_{i}$ still satisfy the property of Cartan subalgebra $[n_{i},n_{j}]=0$.

Using Cartan subalgebra local bases, the density matrix of eq.(\ref{rhod}) is expressed as(Appendix A)
\begin{equation}\label{rhocartan}
\rho=U\rho_{d}U^{\dag}=\frac{1}{N}(I_{N}+\sqrt{2N(N-1)}u^{i}n_{i}),
\end{equation}
where the components $u^{i}(i=1,2,\cdots,N-1)$ are well determined by $a^{n}$ as
\begin{equation}\label{ni}
u^{i}=\sqrt{\frac{N}{N-1}}\frac{1}{\sqrt{i(i+1)}}(\sum_{k=1}^{i}a^{k}-ia^{i+1}).
\end{equation}
So the density matrix of the $N$-level quantum system is parameterized by using $su(N)$ Cartan subalgebra local bases as given in eq.(\ref{rhocartan}) and eq.(\ref{ni}) i.e. $u\equiv u^{i}n_{i}$.

Each local basis $n_{i}$ is a general $su(N)$ Lie algebra vector in parameter space, its covariant derivative is defined as
\begin{equation}\label{Covariant}
  D_{\mu}n_{i}=\partial_{\mu}n_{i}-ig[A_{\mu},n_{i}],
\end{equation}
where $g$ is coupling constant and $A_{\mu}\equiv A_{\mu}^{a}T_{a}$ is $SU(N)$ gauge potential, it is also a Lie algebra vector. We can project $A_{\mu}$ to the directions of local bases of Cartan subalgebra and other directions perpendicular to these local bases\cite{DuanZhang,LiuDuanZhang2005SkySUN,DuanRen}
\begin{equation}\label{Pro}
A_{\mu}=(A_{\mu},n_{i})n_{i}+[[A_{\mu},n_{i}],n_{i}].
\end{equation}
On the other hand, using the definition of $n_{i}$, it can be easily proved that
 \begin{equation}\label{Amu}
\partial_{\mu}n_{i}=[\partial_{\mu}UU^{\dag},n_{i}],
\end{equation}
which means that
 \begin{equation}\label{Amu1}
D_{\mu}n_{i}=\partial_{\mu}n_{i}-ig[K_{\mu},n_{i}]=0,
\end{equation}
where $K_{\mu}\equiv\frac{1}{ig}\partial_{\mu}UU^{\dag}$ and we can verify that it is just the flat connection of $SU(N)$.

Analogous to the 't Hooft $SU(2)$ gauge-invariant electromagnetic tensor\cite{Hooft1974} of magnetic monopoles, Duan and Zhang use the gauge potential decomposition theory to define the extended $SU(N)$ gauge-invariant
 electromagnetic tensor as\cite{DuanZhang,LiuDuanZhang2005SkySUN,DuanRen}
\begin{equation}\label{fmunui1}
f_{\mu\nu}^{i}=(F_{\mu\nu},n_{i})-\frac{1}{ig}(n_{i},[D_{\mu}n_{k},D_{\nu}n_{k}]),
\end{equation}
where $F_{\mu\nu}=\partial_{\mu}A_{\nu}-\partial_{\nu}A_{\mu}-ig[A_{\mu},A_{\nu}]$ is $SU(N)$ gauge field strength. Using eq.(\ref{Pro}) one can prove(Appendix B)
\begin{equation}\label{fmunui2}
f_{\mu\nu}^{i}=\partial_{\mu}A_{\nu}^{i}-\partial_{\nu}A_{\mu}^{i}-\frac{1}{ig}(n_{i},[\partial_{\mu}n_{k},\partial_{\nu}n_{k}]),
\end{equation}
where $A_{\mu}^{i}\equiv(A_{\mu},n_{i})$. When we take $K_{\mu}$ as $A_{\mu}$, the $f_{\mu\nu}^{i}$ in eq.(\ref{fmunui1}) equals $0$ because $D_{\mu}n_{i}=0$ and $F_{\mu\nu}(K)=0$. So according to eq.(\ref{fmunui2}), we obtain
\begin{equation}\label{Wu-Yangcurvature}
K_{\mu\nu}^{i}\equiv\partial_{\mu}a_{\nu}^{i}-\partial_{\nu}a_{\mu}^{i}=\frac{1}{ig}(n_{i},[\partial_{\mu}n_{k},\partial_{\nu}n_{k}]),
\end{equation}
where
 \begin{equation}\label{WuYangPot}
   a_{\mu}^{i}\equiv(K_{\mu},n_{i})=\frac{1}{ig}(\partial_{\mu}UU^{\dag},n_{i})\;\;\;,\;\;\;(i=1,2,\cdots,N-1)
 \end{equation}
is named as $i$-th Wu-Yang potential which is used to describe magnetic monopole\cite{Wu-Yang1}.

Now, thanks to the Cartan subalgebra local bases parametrization of density matrix $\rho$, the physical meanings of these Wu-Yang potentials for a general $N$-level quantum system are clear. At first, it is a remarkable feature that the Wu-Yang potentials in eq.(\ref{WuYangPot}) are completely expressed by local gauge transformations $U(x)$. For $N$-level quantum system, the unitary evolution of density matrix in parameter space is completely described by $U(x)\in SU(N)$, so these Wu-Yang potentials must be produced in any unitary evolution path. Moreover, the parallel transport character for $n_{i}$ defines a parallel transport of density matrix state for this unitary evolution, however there are $(N-1)$ Wu-Yang potentials seen as emergent gauge fields that always exist in this parallel transport state.

On the other hand, $a_{\mu}^{i}$  is the projection of $SU(N)$ flat connection $K_{\mu}\equiv\frac{1}{ig}\partial_{\mu}UU^{\dag}$ on the $i$-th Cartan subalgebra local basis $n_{i}$ and $K_{\mu\nu}^{i}$ is Wu-Yang curvature tensor corresponding to $i$-th Wu-Yang potential $a_{\mu}^{i}$. $K_{\mu\nu}^{i}$ is a $U(1)$ magnetic monopole electromagnetic field tensor, so for $N$-level pure or mixed quantum system, its emergent gauge fields are magnetic monopole electromagnetic fields, but note that we put no constraints on the dimension of parameter space now. The definition of Wu-Yang potentials and Wu-Yang curvature tensors provide us a method for the calculation of emergent electromagnetic fields, it is independent on state and gauge condition but only depend on local gauge transformation $U(x)$. In the following, we mainly use this method to analyze the magnetic monopole property of magnetic skyrmions and the relation between Wu-Yang potentials and Berry connection.

\section{The topological charge of 2-level quantum system}
In this section, we analyze the topological charge of a general 2-level quantum system. For $SU(2)$ case, $\tau _a = \frac{1}{2} \sigma _a(a=1,2,3)$ are generators of $SU(2)$ Lie group, where $\sigma_{a}$ is Pauli matrix. Its Cartan subalgebra index $i$ only equals to $3$, so the $SU(2)$ Wu-Yang potential from eq.(\ref{WuYangPot}) becomes $ a _{\mu}^{3}=\frac{1}{iq_{e}}(\partial_{\mu}UU^{\dag},n_{3})$, where $U\in SU(2)$ and $g$ is substituted with an emergent charge $q_{e}$, which will be explained later. The Cartan subalgebra indices $i,k$ in eq.(\ref{Wu-Yangcurvature}) also should be $3$, so the Wu-Yang curvature tensor becomes
 \begin{equation}\label{WuYangsu(2)}
  K_{\mu\nu}^{3}=\partial_{\mu}a_{\nu}^{3}-\partial_{\nu}a_{\mu}^{3}=\frac{1}{iq_{e}}(n_{3},[\partial_{\mu}n_{3},\partial_{\nu}n_{3}]).
 \end{equation}

 Because $n_{3}$ is a $su(2)$ Lie algebra vector, so we can expand it as $n_{3}=n_{3}^{a}\tau_{a}$, where $n^{a}_{3}\;  (a=1,2,3)$ are its components. Using the commutation relation of Pauli matrices and the definition of the inner product of Lie algebra,
 \begin{eqnarray} \label{Kmunu}
 K_{\mu\nu}^{3}&=&\frac{1}{iq_{e}}(n_{3}^{a}\tau _{a},[\partial_{\mu}n_{3}^{b}\tau _{b},\partial_{\nu}n_{3}^{c}\tau _{c}])\nonumber \\
 &=&\frac{1}{iq_{e}}n_{3}^{a}\partial_{\mu}n_{3}^{b}\partial_{\nu}n_{3}^{c}(\tau _{a},[\tau _{b},\tau _{c}]) \nonumber \\
 &=&\frac{1}{iq_{e}}n_{3}^{a}\partial_{\mu}n_{3}^{b}\partial_{\nu}n_{3}^{c}i\epsilon_{bcd} (\tau _{a},\tau _{d})\nonumber\\  &=&\frac{1}{iq_{e}}n_{3}^{a}\partial_{\mu}n_{3}^{b}\partial_{\nu}n_{3}^{c}2i\epsilon_{bcd}Tr(\tau _{a}\tau _{d})\nonumber\\
 &=&\frac{1}{iq_{e}}n_{3}^{a}\partial_{\mu}n_{3}^{b}\partial_{\nu}n_{3}^{c}2i\epsilon_{bcd}(\frac{1}{2}\delta_{ad}) \nonumber\\
 &=&\frac{1}{q_{e}}\varepsilon_{abc}n_{3}^{a}\partial_{\mu}n_{3}^{b}\partial_{\nu}n_{3}^{c}.
\end{eqnarray}
So the magnetic monopole charge $G$ of $SU(2)$ Wu-Yang potential in $\mathds{R}^{3}$ is\cite{DuanGe1979}
\begin{equation}\label{chaG}
   G=\int_{S^{2} }\frac{1}{2q_{e}}\varepsilon_{abc}n_{3}^{a}\partial_{\mu}n_{3}^{b}\partial_{\nu}n_{3}^{c}dx^{\mu}\wedge dx^{\nu}\,\,\,,\,\,\,(\mu,\nu=1,2).
\end{equation}
which has a electromagnetic field distribution corresponding to its positive or negative charge. When we consider $S^{2}$ as a single-point compactification of a two-dimensional disk $\mathds{D}_{2}$ with a uniform field on the boundary as $\Sigma\equiv\mathds{D}_{2}\cup\{\infty\}\cong S^{2}$,
\begin{equation} \label{MM}
   G=\int_{\Sigma }\frac{1}{2q_{e}}\varepsilon_{abc}n_{3}^{a}\partial_{\mu}n_{3}^{b}\partial_{\nu}n_{3}^{c}dx^{\mu}\wedge dx^{\nu}\,\,\,,\,\,\,(\mu,\nu=1,2).
\end{equation}
Then $G$ corresponds to a quantization magnetic flux in $\Sigma$.

Furthermore, we can see that the Wu-Yang magnetic monopole must have an $n_{3}$ field distribution. Considering $n_{3}$ comes from $ U\tau _{3}U^{\dag}$, $n_{3}$ can be viewed as spin operator in magnetic material and its direction is determined by local gauge transformation $U$. So far, physicists still have not found any real magnetic monopole, but eq.(\ref{MM}) provides us one clue to explore quasi-magnetic monopole in magnetic system. In fact, this configuration of eq.(\ref{MM}) appears in many different materials\cite{Han}. In the following section, we will discuss the physical realization of quasi-magnetic monopole in $(2+1)$-dimensional magnetic material, in which the magnetization configuration can be precisely manipulated in laboratory.

\section{Magnetic skyrmions in two-dimansional materials as quasi-magnetic monopoles}

In this section, using the Cartan subalgebra local basis method introduced in the previous section, we discuss the $SU(2)$ Wu-Yang magnetic monopole charge of magnetic material. In a $2$-dimensional base manifold $\mathcal{M}$ of magnetic material, a smooth magnetization configuration $\vec{M}=\vec{M}(\bm{r},t)$ is a cross section of associative vector bundle $E(\mathcal{M},su(2),SU(2),P)$ and its unit magnetization vector $\vec{m}(\bm{r},t)=\vec{M}(\bm{r},t)/|\vec{M}(\bm{r},t)|$ correspond to a unit $su(2)$ Lie algebra vector $\vec{m}(\bm{r},t) =m^{a}(\bm{r},t)\tau_{a}$ with $\sum_{a}m^{a}(\bm{r},t)m^{a}(\bm{r},t)=1(a=1,2,3)$. We know that $su(2)$ Cartan subalgebra is $\tau _{3}$, and its corresponding local basis is $ n _{3}(\bm{r},t)=U(\bm{r},t)\tau _{3}U^{\dag}(\bm{r},t)$, where $ n _{3}(\bm{r},t)$ remains a $su(2)$ Lie algebra vector and can be expressed as
 \begin{equation}\label{n3com}
  n_{3}(\bm{r},t)=n^{a}_{3}(\bm{r},t)\tau _{a}.
 \end{equation}
From this point of view, we assume
\begin{equation}\label{nm}
 \vec{m}(\bm{r},t)\equiv n_{3}(\bm{r},t),
\end{equation}
that is to say, we describe the local unit magnetization as a $su(2)$ Cartan subalgebra local basis. After replacing the local magnetization $\vec{m}$ with the $su(2)$ Cartan subalgebra local basis $n_{3}$, the $SU(2)$ Wu-Yang potential we discussed in previous section will has a real physical effect that we can detect it in magnetic materials, and we can use it to explain the theoretical origin of the emergent electromagnetic field and Berry phase phenomenal produced in magnetic materials.

According to eq.(\ref{nm}) and the replacing $m^{a}$  with $n_{3}^{a}$, the Wu-Yang curvature tensor $K_{\mu\nu}^{3}$ of $(2+1)$-dimensional magnetic material can be expressed as
     \begin{equation}\label{Kmunum}
      K_{\mu\nu}^{3}=\frac{1}{q_{e}}\varepsilon_{abc}m^{a}\partial_{\mu}m^{b}\partial_{\nu}m^{c} \,\,\,,\,\,\,(\mu,\nu=1,2).
     \end{equation}
According to eq.(\ref{MM}), it is obvious that the Wu-Yang magnetic monopole charge can be expressed as
\begin{equation}\label{MM1}
   G=\int_{\Sigma }\frac{1}{2 q_{e}}\vec{m}\cdot (\partial_{\mu} \vec{m} \times  \partial_{\nu} \vec{m})dx^{\mu}\wedge dx^{\nu}\,\,\,,\,\,\,(\mu,\nu=1,2).
\end{equation}
Many theoretical and experimental studies have proved\cite{skyremg, Han} that skyrmions in $(2+1)$-dimensional magnetic materials have the following structure
\begin{equation}\label{MM2}
   S=\frac{1}{8\pi}\int_{\Sigma }\vec{m}\cdot (\partial_{\mu} \vec{m} \times  \partial_{\nu} \vec{m})dx^{\mu}\wedge dx^{\nu}\,\,\,,\,\,\,\,(\mu,\nu=1,2),
\end{equation}
where $\vec{m}$ is unit magnetization and the value of single skyrmion charge $S$ is $+1$ or $-1$. Now by comparing eq.(\ref{MM1}) and eq.(\ref{MM2}), we find
\begin{equation}\label{KSRelation2}
  S=\frac{q_{e}}{4\pi}G.
\end{equation}
that is to say, a skyrmion number $S$ always correspond to a Wu-Yang magnetic monopole charge $G$, and $q_{e}$ is the emergent electric charge of magnetic skyrmions which satisfies the non-Abelian analog of Dirac quantization condition\cite{Shnir} i.e. eq.(\ref{KSRelation2}). The magnetization configuration itself can be regarded as a quasi-magnetic monopole.

The magnetization configuration of magnetic skyrmions originates from the competition between exchange interaction and DM interaction in a external magnetic field\cite{DM, RohartThiaville}. As a topological protect quasi-particle with many advantages for technological spintronic applications\cite{skyrapp, skyappl}, now magnetic skyrmion is the active research subject of theoretical and experimental. The experiment researchings like magnetic skyrmion topological Hall effect\cite{skyrhall} $et.$ verified that skyrmion has emergent electromagnetic field\cite{skyremg}. Theoretically, the emergent electromagnetic field of magnetic skyrmion originates from a Berry phase, it is obtained by an adiabatic process. However the Wu-Yang potential can be expressed as a general gauge transformation, in spite of adiabatic or not, so we think that the $SU(2)$ Wu-Yang potential is a theoretical origin of spin Berry phase, we will further explain it in the following section.
\section{Berry connection in gauge transformation viewpoint}
In the preceding section, we find the relation between $SU(2)$ Wu-Yang monopole charge and skyrmion number. In this section, we connect our Cartan subalgebra local bases method to spin Berry connection, and then we discuss the general relation between $SU(N)$ Wu-Yang potentials and Berry connection of pure or mixed state.

In ferromagnetic materials, the spin Berry connection is defined as the overlap of the neighbor spin coherent states\cite{Tatara}. Now we express the spin Berry connection of a $(2+1)$-dimensional magnetization configuration $\vec{m}(\bm{r},t)$ using local $SU(2)$ gauge transformation, and discuss the relation between Wu-Yang pontential and spin Berry connection. At first, we assume a uniform magnetization configuration $\vec{m}_{0}(\bm{r})=\tau_{3}$, so the local gauge transformation related $\vec{m}_{0}(\bm{r})$ to $\vec{m}(\bm{r},t)$ is
\begin{equation}\label{U}
U(\bm{r},t)=\exp(-i\frac{\theta}{2}\vec{\sigma}\cdot\vec{\ell})=\cos\frac{\theta}{2}I-i \sin\frac{\theta}{2}\vec{\sigma}\cdot\vec{\ell},\\
\end{equation}
explicitly
\begin{equation}\label{vecm}
\vec{m}(\bm{r},t)=U(\bm{r},t)\vec{m}_{0}(\bm{r})U^{\dag}(\bm{r},t).\\
\end{equation}
where $\vec{l}$ is the rotation axis
\begin{equation}
\vec{\ell}=(-\sin\phi,\cos\phi,0),
\end{equation}
and $\theta(\bm{r},t),\phi(\bm{r},t)$ are polar angle and azimuth of $\vec{m}(\bm{r},t)$. The spin coherent states corresponding to $\vec{m}(\bm{r},t)$ are
\begin{eqnarray}
|\vec{m}(\bm{r},t)\rangle_{\uparrow}&=&{
  \left( \begin{array}{cc}
\cos\frac{\theta}{2} \\
e^{i\phi}\sin\frac{\theta}{2}
\end{array}
\right )}  \;\;,  \nonumber \\
|\vec{m}(\bm{r},t)\rangle_{\downarrow} &=&{
  \left( \begin{array}{cc}
-e^{-i\phi}\sin\frac{\theta}{2} \\
\cos\frac{\theta}{2}
\end{array}
\right )}.
\end{eqnarray}
So the gauge transformation from $|\vec{m}_{0}(\bm{r})\rangle$ to $|\vec{m}(\bm{r},t)\rangle$ is expressed as
\begin{equation}\label{eigm}
|\vec{m}(\bm{r},t)\rangle_{s}=U(\bm{r},t)|\vec{m}_{0}(\bm{r})\rangle_{s}\,\,\,,\,\,\,\,s=\uparrow \;\mbox{or}\;\downarrow,
\end{equation}
where $|\vec{m}_{0}(\bm{r})\rangle_{\uparrow}={
  \left( \begin{array}{cc}
1 \\
0
\end{array}
\right )}$
and
$|\vec{m}_{0}(\bm{r})\rangle_{\downarrow}={
  \left( \begin{array}{cc}
0 \\
1
\end{array}
\right )}$. It is well known that the matrix form of $U$ is
\begin{equation}\label{matU}
  U=U^{0}I+iU^{a}\sigma_{a}={
  \left( \begin{array}{cc}
U^{0}+iU^{3} & iU^{1}+U^{2} \\
iU^{1}-U^{2} & U^{0}-iU^{3}
\end{array}
\right )},
\end{equation}
with
\begin{eqnarray}
    U^{0}&=&\cos\frac{\theta}{2},\;\;U^{1}=\sin\frac{\theta}{2}\sin\phi, \nonumber \\
    U^{2}&=&-\sin\frac{\theta}{2}\cos\phi,\;\;U^{3}=0.
\end{eqnarray}

Moreover, we identify this gauge transformation eq.({\ref{U}}) with a unitary time-evolution operator which evolve the initial state $|\vec{m}_{0}(\bm{r})\rangle$ to $|\vec{m}(\bm{r},t)\rangle$ with a proper Hamiltonian $h(t)$, and it is expressed as\cite{Niu}
\begin{equation}
U=\textrm{T}(e^{\frac{1}{i}\int_{0}^{t}dt^{\prime}h(t^{\prime})}),
\end{equation}
where $\textrm{T}$ is the time-ordering operator.
The adiabatic approximation condition in time-evolution process define a process without jumping from a given eigenvalue to another eigenvalue, that is
\begin{eqnarray}\label{adia}
_{s}\langle \vec{m}(\bm{r},t)|\vec{m}(\bm{r^{\prime}},t^{\prime})\rangle_{s^{\prime}}=0\,\, , \,\,\,\,  \mbox{for all}\,\, s\neq s^{\prime}.
\end{eqnarray}
Under this approximation conditon, the spin Berry connection\cite{Tatara,Wen} is calculated as
\begin{eqnarray}\label{Aup}
  \mathcal{A}_{\mu\uparrow}&=&-i_{\uparrow}\langle \vec{m}(\bm{r},t)|\vec{m}(\bm{r^{\prime}},t^{\prime})\rangle_{\uparrow} \nonumber\\
&=&-i_{\uparrow}\langle \vec{m}_{0}(\bm{r})|U^{\dag}(\bm{r},t)\partial_{\mu}U(\bm{r},t)|\vec{m}_{0}(\bm{r^{\prime}})\rangle_{\uparrow} \nonumber\\
&=&-i(\partial_{\mu}(U^{0}+iU^{3})(U^{0}-iU^{3}) \nonumber\\
&& +\partial_{\mu}(iU^{1}-U^{2})(-iU^{1}-U^{2}))\nonumber\\
&=&\sin^{2}\frac{\theta}{2}\partial_{\mu}\phi\nonumber\\
&=&\frac{1}{2}(1-\cos\theta)\partial_{\mu}\phi.
\end{eqnarray}
and
\begin{eqnarray}\label{Adown}
  \mathcal{A}_{\mu\downarrow} &=& -i_\downarrow\langle \vec{m}(\bm{r},t)|\vec{m}(\bm{r^{\prime}},t^{\prime})\rangle_\downarrow \nonumber\\
&=&-i_\downarrow\langle \vec{m}_{0}(\bm{r})|U^{\dag}(\bm{r},t)\partial_{\mu}U(\bm{r},t)|\vec{m}_{0}(\bm{r^{\prime}})\rangle_\downarrow \nonumber\\
&=&i(\partial_{\mu}(iU^{1}+U^{2})(iU^{1}-U^{2}) \nonumber\\
&& +\partial_{\mu}(U^{0}-iU^{3})(-U^{0}-iU^{3}))\nonumber\\
&=&-\sin^{2}\frac{\theta}{2}\partial_{\mu}\phi\nonumber\\
&=&\frac{1}{2}(\cos\theta-1)\partial_{\mu}\phi.
\end{eqnarray}
So we define a $SU(2)$ Berry connection matrix as
\begin{equation}\label{Amat}
  A_{\mu}\equiv\textrm{Diag}(\mathcal{A}_{\mu\uparrow},\mathcal{A}_{\mu\downarrow}).
\end{equation}

On the other hand, for the same gauge transformation $U(\bm{r},t)$, the $SU(2)$ Wu-Yang pontential can be calculated as
\begin{eqnarray}\label{WuYang}
  a_{\mu}^{3}&=&(K_{\mu},n_{3}) \nonumber\\
  &=&\frac{1}{iq_{e}}(\partial_{\mu}UU^{\dag},U\tau_{3}U^{\dag})\nonumber\\
  &=&\frac{2}{iq_{e}}\textrm{Tr}(\partial_{\mu}UU^{\dag}U\tau_{3}U^{\dag}) \nonumber\\
  &=&\frac{2}{iq_{e}}\textrm{Tr}(\partial_{\mu}U\tau_{3}U^{\dag})\nonumber\\
  &=&\frac{1}{iq_{e}}\textrm{Tr}\left[
    \left( \begin{array}{cc}
\partial_{\mu}(U^{0}+iU^{3}) & \partial_{\mu}(iU^{1}+U^{2}) \nonumber\\
\partial_{\mu}(iU^{1}-U^{2}) & \partial_{\mu}(U^{0}-iU^{3})
\end{array}
\right) \right. \\
&&\;\;\;\;\;\; \left. \left( \begin{array}{cc}
U^{0}-iU^{3} & -iU^{1}-U^{2} \nonumber\\
iU^{1}-U^{2} & -U^{0}-iU^{3}
\end{array}
\right) \right]\nonumber\\
&=&\frac{1}{iq_{e}}[ \partial_{\mu}(U^{0}+iU^{3})(U^{0}-iU^{3})\nonumber\\
&&+\partial_{\mu}(iU^{1}+U^{2})(iU^{1}-U^{2})\nonumber\\
&&+\partial_{\mu}(iU^{1}-U^{2})(-iU^{1}-U^{2})\nonumber\\
&&+ \partial_{\mu}(U^{0}-iU^{3})(-U^{0}-iU^{3})]\nonumber\\
&=&\frac{2}{q_{e}}\sin^{2}\frac{\theta}{2}\partial_{\mu}\phi\nonumber\\
&=&\frac{1}{q_{e}}(1-\cos\theta)\partial_{\mu}\phi.
\end{eqnarray}
So obviously
\begin{equation}\label{Aa}
a_{\mu}^{3}=\frac{2}{q_{e}}\mathcal{A}_{\mu\uparrow}=-\frac{2}{q_{e}}\mathcal{A}_{\mu\downarrow}\,\,\,\textrm{and}\,\,\,
A_{\mu}=q_{e}a_{\mu}^{3}\tau_{3}.
\end{equation}
So the spin Berry connection of a smooth magnetization configuration is proportional to Wu-Yang potential with a factor $\pm\frac{2}{q_{e}}$. Particularly, to a static skyrmion  $\vec{m}(\bm{r})$, a real space spin Berry connection serves as emergent gauge field which originates from $SU(2)$ Wu-Yang potential exactly.

Now we study the $SU(N)$ Berry connection and Wu-Yang poteneials in adiabatic unitary evolution. Similar to the case of $SU(2)$ magnetization, the uniform field in parameter space that we need is\cite{mixgeo, mixgeo2}
\begin{equation}
u_{0}(x)\equiv\rho_{d}=a^{n}\rho_{n}.
\end{equation}
There are $N$ orthogonal eigenstates for $u_{0}(x)$
\begin{eqnarray}
&&|u_{0}(x)\rangle_{1}={
  \left( \begin{array}{cc}
1 \\
0\\
\vdots\\
0
\end{array}
\right )}\,\,\,,\,\,\,
|u_{0}(x)\rangle_{2}={
  \left( \begin{array}{cc}
0 \\
1\\
\vdots\\
0
\end{array}
\right )}\,\,\,,\,\,\,\cdots\,\,\,,\nonumber\\
&&
|u_{0}(x)\rangle_{N-1}={
  \left( \begin{array}{cc}
0\\
\vdots\\
1\\
0
\end{array}
\right )}\,\,\,,\,\,\,
|u_{0}(x)\rangle_{N}={
  \left( \begin{array}{cc}
0\\
0\\
\vdots\\
1
\end{array}
\right )}.
\end{eqnarray}
For an explicit $U(x,t)\in SU(N)$, the unitary evolution path is
\begin{equation}
u(x,t)=U(x,t)u_{0}(x)U^{\dag}(x,t),
\end{equation}
and the orthogonal eigenstates for $u(x,t)$ are
\begin{equation}
|u(x,t)\rangle_{n}=U(x,t)|u_{0}(x)\rangle_{n}.
\end{equation}
Under the adiabatic approximation conditon eq.(\ref{adia}), the Berry
connections for $SU(N)$ are calculated as
\begin{eqnarray}
\mathcal{A}_{\mu n}&=&-i_{n}\langle m(x,t)|m(x^{\prime},t^{\prime})\rangle_{n} \nonumber\\
&=&-i_{n}\langle m_{0}(x)|U^{\dag}(x,t)\partial_{\mu}U(x,t)|m_{0}(x^{\prime})\rangle_{n}\nonumber\\
&=&-\frac{i}{2}(U^{\dag}(x,t)\partial_{\mu}U(x,t),\rho_{n}).
\end{eqnarray}
Now the $SU(N)$ Berry connection matrix can be defined as
\begin{equation}
  A_{\mu}=\textrm{Diag}(\mathcal{A}_{\mu1},\mathcal{A}_{\mu2},\cdots,\mathcal{A}_{\mu N}).
\end{equation}

The weighted average of $A_{\mu}$ for pure or mixed state is an average of $\mathcal{A}_{\mu n}$ with probability $a^{n}$\cite{mixgeo},
\begin{eqnarray}\label{talBerry}
\mathcal{A}_{\mu}=-\frac{i}{2}(U^{\dag}(x,t)\partial_{\mu}U(x,t),\rho_{d}).
\end{eqnarray}
Reversing eq.(\ref{rhocartan}) we obtain
\begin{eqnarray}
\rho_{d}=U^{\dag}\rho U,
\end{eqnarray}
then
\begin{eqnarray}\label{talBerry}
\mathcal{A}_{\mu}&=&-\frac{i}{2}(\partial_{\mu}U(x,t)U^{\dag}(x,t),\rho)\nonumber\\
&=&\frac{g}{2}(K_{\mu},\rho).
\end{eqnarray}
So the explicit meaning of the weighted average of $A_{\mu}$ is the expectation value of flat connection respect to the final density matrix state in adiabatic unitary evolution.

Comparing eq.(\ref{talBerry}) with eq.(\ref{WuYangPot}) of $SU(N)$ Wu-Yang potentials, using eq.(\ref{rhocartan}), the exact relation between Berry connection of weighted average for pure or mixed state and Wu-Yang potentials is obtained
\begin{eqnarray}
\mathcal{A}_{\mu}&=&\frac{g}{2}(K_{\mu},\rho)\nonumber\\
&=&\frac{g}{2N}(K_{\mu},(I_{N}+\sqrt{2N(N-1)}u^{i}n_{i}))\nonumber\\
&=&\frac{g}{2}\sqrt{\frac{2(N-1)}{N}}u^{i}(K_{\mu},n_{i})\nonumber\\
&=&\frac{g}{2}\sqrt{\frac{2(N-1)}{N}}u^{i}a_{\mu}^{i},
\end{eqnarray}
where the $u^{i}$ is expressed as $a^{n}$ by eq.(\ref{ni}).
In the case of $SU(2)$ magnetic skyrmion, $N=2$ and the unit magnetization vector correspond to pure state i.e. $a^{1}=1,a^{2}=0$ or $a^{1}=0,a^{2}=1$, so using eq.(\ref{ni}) we obtain $u^{1}=a^{1}=1$ or $u^{1}=a^{2}=-1$, that is to say
\begin{eqnarray}
\mathcal{A}_{\mu}=\frac{g}{2}a_{\mu}^{3}\,\,\, \textrm{or}\,\,\, \mathcal{A}_{\mu}=-\frac{g}{2}a_{\mu}^{3},
\end{eqnarray}
that is exactly the eq.(\ref{Aa}).

To summarize the discussion above, the methods of creating or deleting magnetic skyrmion in magnetic material\cite{skyrapp} are equivalent to creating or deleting quasi-magnetic monopole by taking $SU(2)$ gauge transformation for the magnetization vector. For $SU(N)$ case, we take the gauge transformation on $\rho_{d}$, then the $(N-1)$ Wu-Yang potentials are theoretically related to $(N-1)$ topological quasi-particles for pure or mixed state, these topological quasi-particles are similar to magnetic skyrmion in $SU(2)$, so the result of $SU(N)$ provide us a new clue for creating topological quasi-particles.
\section{Conclusion}
Since magnetic skyrmions were observed in ferromagnetic materials\cite{skyrob1, skyrob2}, various novel properties of magnetic skyrmion have been detected which mainly owing to its topological characteristics. Usually, the emergent electromagnetic field of magnetic skyrmions is considered to be a useful tool for understanding these properties\cite{skyremg}. To obtain this emergent electromagnetic field from the magnetization configuration of skyrmion itself, using the method of gauge transformation, we take the local magnetization in magnetic material as a $su(2)$ Cartan subalgebra local basis. Then we verified that the $SU(2)$ Wu-Yang potential equals to the adiabatical Berry connection multiplied by a factor $\pm\frac{2}{q_{e}}$, so the same algebraic structure between magnetic skyrmion and Wu-Yang monopole suggest that the magnetic skyrmion is quasi-magnetic monopole.

Moreover, by extending the discussion from $SU(2)$ case to $SU(N)$, we find that the magnetization is  naturally replaced by density operator $\rho$. And then, by applying the $su(N)$ Cartan subalgebra parametrization method to this density operator, we obtain $(N-1)$ Wu-Yang potentials of a general $N$-level quantum system. After considering the adiabatic approximation, we obtain $N$ Berry connections for the adiabatic unitary evolution, the weighted average for these Berry connections is exactly the expectation value of flat connection with respect to the final state of adiabatic unitary evolution. Finally, we find that the $(N-1)$ Wu-Yang potentials can be used to calculate the weighted average of Berry connections, it is a new representation different from ref.\cite{mixgeo}. Analogous to $SU(2)$ magnetic skyrmions, the result of $SU(N)$ case provide us a new clue for creating topological quasi-particles.

\section{Appendix A}
In our conventions of normalization relation eq.(\ref{nor}) and commutate and anticommutate relations eq.(\ref{com}) for $su(N)$ Lie algebra, the generators $H_{i}(i=1,2,\cdots,N-1)$ of $su(N)$ Cartan subalgebra are
\begin{eqnarray}
H_{i}=\frac{1}{\sqrt{2i(i+1)}}\textrm{Diag}(1,1,\cdots,1,-i,0,\cdots,0),
\end{eqnarray}
where $-i$ is the $(i+1)$-th diagonal element. The $\rho_{d}$ in eq.(\ref{rhod}) is
\begin{eqnarray}
\rho_{d}\equiv a^{n}\rho_{n}=\textrm{Diag}(a^{1},a^{2},\cdots,a^{N}),
\end{eqnarray}
with $\rho_{n}=\textrm{Diag}(0,0,\cdots,1,0,\cdots,0)$, where $1$ is the $n$-th diagonal element. A recursive relation between $\rho_{n}(n=1,2,\cdots,N)$ and $H_{i}(i=1,2,\cdots,N-1)$ is
\begin{eqnarray}
\rho_{n+1}-\rho_{n}=\sqrt{\frac{2(n-1)}{n}}H_{n-1}-\sqrt{\frac{2(n+1)}{n}}H_{n}.
\end{eqnarray}
After summation, the $\rho_{n}$ is expressed as
\begin{eqnarray}
\rho_{n}=&&-\sqrt{\frac{2n}{n-1}}H_{n-1}+(\sqrt{\frac{2(n-2)}{n-1}}-\sqrt{\frac{2(n-1)}{n-2}})H_{n-2}\nonumber\\
&&+\cdots+(1-2)H_{1}+\rho_{1}.
\end{eqnarray}
We calculate the $\rho_{1}$ as
\begin{equation}
\rho_{1}=\frac{1}{N}I_{N}+\sum_{k=1}^{N-1}\sqrt{\frac{2}{k(k+1)}}H_{k}.
\end{equation}
Then
\begin{eqnarray}
\rho_{d}&=&a^{n}\rho_{n}\nonumber\\
&=&\sum_{i=1}^{N-1}[(\sqrt{\frac{2i}{i+1}}-\sqrt{\frac{2(i+1)}{i}})(a^{N}+a^{N-1}+\cdots\nonumber\\
&&~+a^{i+2})-\sqrt{\frac{2(i+1)}{i}}a^{i+1}]H_{i}+\rho_{1}\nonumber\\
&=&\sum_{i=1}^{N-1}[(-\sqrt{\frac{2}{i(i+1)}})(1-(a^{1}+a^{2}+\cdots+a^{i+1}))\nonumber\\
&&-(i+1)\sqrt{\frac{2}{i(i+1)}}a^{i+1}]H_{i}+\rho_{1}\nonumber\\
&=&\sum_{i=1}^{N-1}\sqrt{\frac{2}{i(i+1)}}[a^{1}+a^{2}+\cdots+a^{i}\nonumber\\
&&-ia^{i+1}-1]H_{i}+\rho_{1}\nonumber\\
&=&\frac{1}{N}I_{N}+\sum_{i=1}^{N-1}\sqrt{\frac{2}{i(i+1)}}[a^{1}+a^{2}+\cdots\nonumber\\
&&+a^{i}-ia^{i+1}]H_{i}.
\end{eqnarray}
So
\begin{equation}
\rho=U\rho_{d}U^{\dag}=\frac{1}{N}(I_{N}+\sqrt{2N(N-1)}u^{i}n_{i}),
\end{equation}
with
\begin{equation}
u^{i}=\sqrt{\frac{N}{N-1}}\frac{1}{\sqrt{i(i+1)}}(\sum_{k=1}^{i}a^{k}-ia^{i+1}).
\end{equation}
\section{Appendix B}
We begin from the projection formula eq.(\ref{Pro})
\begin{eqnarray}\label{pro}
A_{\mu}=(A_{\mu},n_{k})+[[A_{\mu},n_{k}],n_{k}].
\end{eqnarray}
The differential of $A_{\mu}$ is
\begin{eqnarray}
\partial_{\mu}A_{\nu}&=&\partial_{\mu}A_{\nu}^{k}n_{k}+A_{\nu}^{k}\partial_{\mu}n_{k}+[\partial_{\mu}[A_{\nu},n_{k}],n_{k}]\nonumber \\
&&+[[A_{\nu},n_{k}],\partial_{\mu}n_{k}].
\end{eqnarray}
and the projection of $\partial_{\mu}A_{\nu}$ to $n_{i}$ is
\begin{eqnarray}\label{pamupanuni}
&&((\partial_{\mu}A_{\nu}-\partial_{\mu}A_{\nu}),n_{i})\nonumber\\
&=&\partial_{\mu}A_{\nu}^{k}(n_{k},n_{i})+A_{\nu}^{k}(\partial_{\mu}n_{k},n_{i}) \nonumber\\
&&+([\partial_{\mu}[A_{\nu},n_{k}],n_{k}],n_{i})+([[A_{\nu},n_{k}],\partial_{\mu}n_{k}],n_{i})\nonumber\\
&&-\partial_{\nu}A_{\mu}^{k}(n_{k},n_{i})-A_{\mu}^{k}(\partial_{\nu}n_{k},n_{i}) \nonumber\\
&&-([\partial_{\nu}[A_{\mu},n_{k}],n_{k}],n_{i})-([[A_{\mu},n_{k}],\partial_{\nu}n_{k}],n_{i})\nonumber\\
&=&([[A_{\nu},n_{k}],\partial_{\mu}n_{k}],n_{i})-([[A_{\mu},n_{k}],\partial_{\nu}n_{k}],n_{i})\nonumber\\
&&+(\partial_{\mu}A_{\nu}^{i}-\partial_{\nu}A_{\mu}^{i}),
\end{eqnarray}
where we used $(n_{i},[n_{k},L])=0$ for any $L\in su(N)$ in the second equal sign.

Any Lie algebra vectors $A,B,C$ satisfy Jacobbi identity
\begin{eqnarray}
[A,[B,C]]+[B,[C,A]]+[C,[A,B]]=0.
\end{eqnarray}
We take $A=[A_{\mu},n_{k}],B=A_{\nu},C=n_{k}$, so the Jacobbi identity is
\begin{eqnarray}
 &&[[A_{\mu},n_{k}],[A_{\nu},n_{k}]]  \nonumber \\
&=&-[A_{\nu},[n_{k},[A_{\mu},n_{k}]]]-[n_{k},[[A_{\mu},n_{k}],A_{\nu}]].
\end{eqnarray}
Then the projection of $[[A_{\mu},n_{k}],[A_{\nu},n_{k}]]$ to $n_{i}$ is
\begin{eqnarray}\label{AmuAnuni}
&& ([[A_{\mu},n_{k}],[A_{\nu},n_{k}]],n_{i}) \nonumber\\
&=&-([A_{\nu},[n_{k},[A_{\mu},n_{k}]]],n_{i})-([n_{k},[[A_{\mu},n_{k}],A_{\nu}]],n_{i})\nonumber\\
&=&-([A_{\nu},[n_{k},[A_{\mu},n_{k}]]],n_{i})\nonumber\\
&=&([A_{\nu},[[A_{\mu},n_{k}],n_{k}]],n_{i}) \nonumber\\
&=&-([[[A_{\mu},n_{k}],n_{k}],A_{\nu}],n_{i}) \nonumber\\
&=&([(A_{\mu}-A_{\mu}^{k}n_{k}),A_{\nu}],n_{i})\nonumber\\
&=&-([A_{\mu},A_{\nu}],n_{i}),
\end{eqnarray}
where we used $(n_{i},[n_{k},L])=0$ for any $L\in su(N)$ in the second equal sign, and we used eq.(\ref{pro}) in the fifth equal sign.
According to
\begin{equation}
D_{\mu}n_{k}=\partial_{\mu}n_{k}-ig[A_{\mu},n_{k}],
\end{equation}
 we obtain
\begin{eqnarray}
&& ([D_{\mu}n_{k},D_{\nu}n_{k}],n_{i}) \\
&=&([(\partial_{\mu}n_{k}-ig[A_{\mu},n_{k}]),(\partial_{\nu}n_{k}-ig[A_{\nu},n_{k}])],n_{i})\nonumber\\
&=&([\partial_{\mu}n_{k},\partial_{\nu}n_{k}],n_{i})+(ig)^{2}([A_{\mu},n_{k}],[A_{\nu},n_{k}],n_{i})\nonumber\\
&&-ig([\partial_{\mu}n_{k},[A_{\nu},n_{k}]],n_{i})-ig([[A_{\mu},n_{k}],\partial_{\nu}n_{k}],n_{i}). \nonumber
\end{eqnarray}
After transposition, we obtain
\begin{eqnarray}
&&([[A_{\nu},n_{k}],\partial_{\mu}n_{k}],n_{i})-([[A_{\mu},n_{k}],\partial_{\nu}n_{k}],n_{i})\nonumber\\
&=&\frac{1}{ig}([D_{\mu}n_{k},D_{\nu}n_{k}],n_{i})-\frac{1}{ig}([\partial_{\mu}n_{k},\partial_{\nu}n_{k}],n_{i}) \nonumber\\
&&-ig([A_{\mu},n_{k}],[A_{\nu},n_{k}],n_{i}))\nonumber\\
&=&\frac{1}{ig}([D_{\mu}n_{k},D_{\nu}n_{k}],n_{i})-\frac{1}{ig}([\partial_{\mu}n_{k},\partial_{\nu}n_{k}],n_{i}) \nonumber\\
&&+ig([A_{\mu},A_{\nu}],n_{i}).
\end{eqnarray}
where we used eq.(\ref{AmuAnuni}) in second equal sign.
After transposition, and using eq.(\ref{pamupanuni}) we obtain
\begin{eqnarray}
f_{\mu\nu}^{i}&\equiv&(F_{\mu\nu},n_{i})-\frac{1}{ig}([D_{\mu}n_{k},D_{\nu}n_{k}],n_{i})\nonumber\\
&=&(\partial_{\mu}A_{\nu}^{i}-\partial_{\nu}A_{\mu}^{i})-\frac{1}{ig}([\partial_{\mu}n_{k},\partial_{\nu}n_{k}],n_{i}),
\end{eqnarray}
where $F_{\mu\nu}=\partial_{\mu}A_{\nu}-\partial_{\nu}A_{\mu}-ig[A_{\mu},A_{\nu}]$.

\end{document}